\documentclass[a4paper]{jpconf}
\usepackage{graphicx}

\newcommand{\beq}{\begin{equation}}
\newcommand{\eeq}{\end{equation}}
\newcommand{\bea}{\begin{eqnarray}}
\newcommand{\eea}{\end{eqnarray}}

\newcommand{\n}{\nu}
\newcommand{\m}{\mu}

\newcommand{\prt}{\partial}

\newcommand{\cD}{{\cal D}}

\newcommand{\cH}{{\cal H}}

\newcommand{\ket}[1]{ \mid\! #1 \rangle}
\newcommand{\bra}[1]{ \langle #1 \!\mid}

\begin{document}
\title{Causal random geometry from stochastic quantization}

\author{J Ambj{\o}rn$^{1,2}$, R Loll$^2$, 
W Westra$^{3}$ and S Zohren$^{4,5}$}

\address{$^1$ Niels Bohr Institute, Copenhagen, Denmark}
\address{$^2$ Institute for Theoretical Physics, Utrecht University, The Netherlands}
\address{$^3$ Department of Mathematics, University of Iceland, Iceland}
\address{$^4$ Mathematical Institute, Leiden University, The Netherlands}
\address{$^5$ Blackett Laboratory, Imperial College London, UK}

\ead{ambjorn@nbi.dk,loll@phys.uu.nl,wwestra@raunvis.hi.is, zohren@math.leidenuniv.nl}

\begin{abstract}
In this short note we review a recently found formulation of two-dimensional causal quantum gravity defined through Causal Dynamical Triangulations and stochastic quantization. This procedure enables one to extract the nonperturbative quantum Hamiltonian of the random surface model including the sum over topologies. Interestingly, the generally fictitious stochastic time corresponds to proper time on the geometries.
\end{abstract}

\section{Introduction}
Two-dimensional random geometry can either be viewed as a simple toy model for a string theory of fundamental particles or as a playground for four-dimensional quantum gravity. The simplification from the string theory point of view is that the two-dimensional random surfaces we discuss are not embedded in an ambient space, only the intrinsic two-dimensional geometry of the string is considered. Although this removes many features of a full blown string theory, it allows us to focus on the signature of the metric on the surface, also called worldsheet, of the propagating string. In traditional approaches to two-dimensional random geometry, matrix models and Liouville theory, the signature of the metric is considered to be Euclidean whereas genuine string theory is presumed to possess Lorentzian signature worldsheets. Causal Dynamical Triangulations (CDT) describes the random geometry of two-dimensional surfaces with Lorentzian signature without using coordinates, hence avoiding complications due to gauge invariance \cite{al}.  The essential difference between causal surfaces and Euclidean surfaces is that the time dependent process where a string splits into two strings is nonsingular in the Euclidean case and singular in the Lorentzian case. This leads to a different assignment of coupling constants and a markedly different continuum theory \cite{cap}. In this short note we review a recent observation \cite{stoch} that the time on the Lorentzian surfaces can be identified with the  ``fifth time'' variable of the stochastic quantization method introduced by Parisi and Wu \cite{parisi}. Remarkably the splitting and merging dynamics of the string with respect to this stochastic time can be described by an \emph{exactly solvable} Schr\"{o}dinger like equation \cite{stoch}! In other words the method of stochastic quantization provides a natural nonperturbative definition of the quantum dynamics of causal random surfaces of arbitrary topology.

\section{Stochastic quantization}

Stochastic quantization was introduced by Parisi and Wu \cite{parisi} in 1981 as a quantization scheme for Euclidean field theories (see \cite{dam} for a standard review and \cite{dijk} for an interesting more recent account). The basic idea is to start off with a statistical physics system which evolves according to a, in general, fictitious time $t$ and to observe that there exists a correspondence between certain correlation functions in the equilibrium statistical physics model at $t\to\infty$ and propagators in the Euclidean field theory. We will outline this correspondence in the simplest case of zero-dimensional Euclidean field theory which is sufficient for our purposes.

Consider a time dependent random variable $X_t$ with initial value $X_0=x_0$ whose time evolution is given by the Langevin equation
\beq\label{eq:langevin}
\frac{dX_t}{dt}=-f(X_t) +\sqrt{2 \Omega} \,\nu(t),\quad f(x)=\frac{\partial S(x)}{\partial x},
\eeq
where $\nu(t)$ is Gaussian white noise of width one and $f(x)$ is a conservative drift. Let us now consider the probability density for $X_t=x$ given that $X_0=x_0$,
\beq
P(x,x_0,t)=E_\nu\{ \delta (x-X_t) | X_0=x_0 \},
\eeq
where $E_\nu$ refers to the expectation value with respect to the Gaussian white noise $\nu(t)$. From the Langevin equation \eref{eq:langevin} one can then derive the so-called Fokker-Planck equation for the probability density
\beq \label{eq:FP}
\frac{\prt}{\prt t} P(x,x_0,t)=\frac{\prt}{\prt x}\left(\frac{\prt S}{\prt x}+\Omega \frac{\prt}{\prt x}  \right)P(x,x_0,t).
\eeq
From this one can directly verify that for $t\to\infty$,
\beq \label{eq:eq}
\lim_{t\to\infty} P(x,x_0,t)=P^{eq}(x)=\frac{1}{Z}e^{-\frac{1}{\Omega}S(x)},\quad Z= \int dx e^{-\frac{1}{\Omega}S(x)}
\eeq
which is the correct measure arising from the Euclidean path integral with classical action $S(x)$ and thus points towards the correspondence with Euclidean field theory. Here $\Omega$ can be see as the analog of $\hbar$ in the quantum mechanical setting. 

In principle one can now use \eref{eq:FP} to extract the effective quantum Hamiltonian of the model with classical action $S(x)$. In the following we use this procedure to deduce the effective quantum Hamiltonian of a model of two-dimensional causal quantum gravity which we introduce in the forthcoming section.

\section{Two-dimensional causal quantum gravity}

The Causal Dynamical Triangulation (CDT) approach to quantum gravity 
aims to define two-dimensional Lorentzian quantum gravity through a nonperturbative path integral over geometries $[g_{\m\n}]$
\beq\label{eq:partition}
Z(\kappa,\lambda)= \int \cD [g_{\m\n}] \; \e^{-S[g_{\m\n}]},
\eeq
where the (Euclidean) Einstein-Hilbert action is given by
\beq \label{eq:action}
S[g_{\m\n}] = -\frac{1}{2\pi G_N} \int \sqrt{\det{g_{\m\n}}}\;R + \lambda \int \sqrt{\det g_{\m\n}}.
\eeq
Here $\lambda$ denotes the cosmological constant, $G_N$ the Newton's constant and $\kappa= \e^{-1/G_N}$ is the dimensionless string coupling. 

In the lattice regularization of CDT the path integral is defined as a continuum limit of a sum over triangulated surfaces (see \cite{al} or the recent review \cite{review}). This is done still with Lorentzian signature and a Wick rotation to Euclidean signature is performed at the level of the individual triangulations. In contrast to the Euclidean model defined through Dynamical Triangulations (DT), the Euclidean path integral of the Lorentzian model defined through CDT includes only geometries with a fixed time-sliced structure disallowing for spatial topology changes.

\begin{figure}[t]
\begin{minipage}{14pc}\vspace{1pc}
\includegraphics[width=12pc]{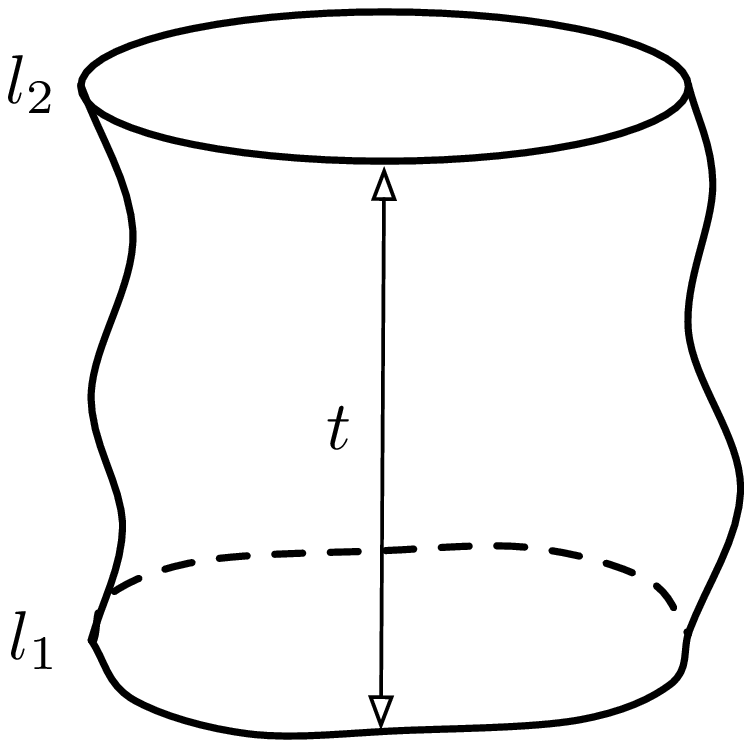}\vspace{2pc}
\caption{\label{fig1} Illustration of the propagator $G_0(l_1,l_2;t)$, describing the amplitude from a fixed spatial boundary of length $l_1$ to a boundary of length $l_2$ in time $t$.}
\end{minipage}\hspace{2pc}%
\begin{minipage}{14pc}
\includegraphics[width=12pc]{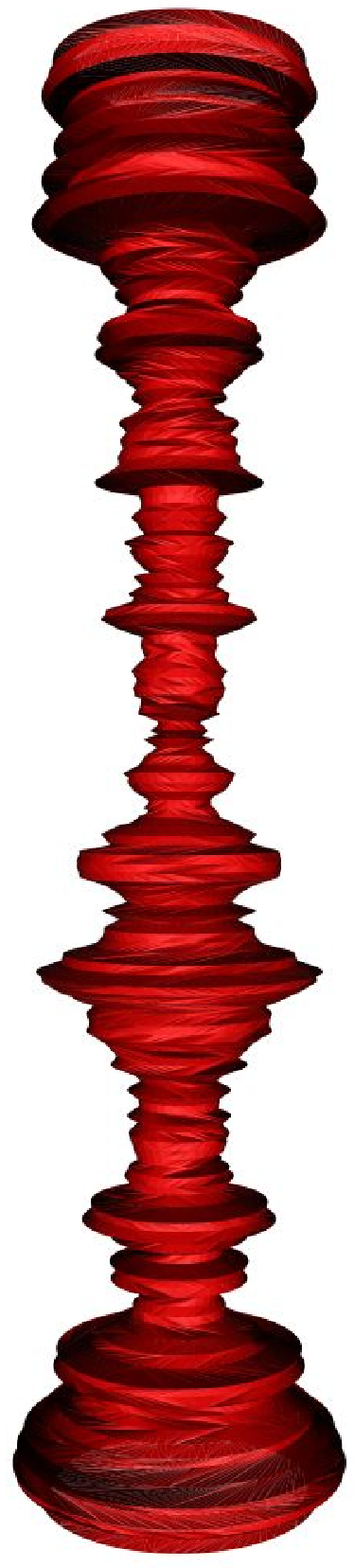}
\caption{\label{fig2} Snapshot from a Monte Carlo simulation of a two-dimensional CDT universe.}
\end{minipage} 
\end{figure}

Let us denote the (genus zero) propagator from an initial marked spatial slice of length $l_1$ to a final unmarked spatial slice of length $l_2$ in proper time $t$ by $G_0(l_1,l_2;t)$ (see Fig.\ \ref{fig1}). Let us for convenience also introduce the Laplace transformed propagator 
\beq
G_0 (x_1,x_2;t) = \int_0^\infty d l_1  \int_0^\infty d l_2
\; \e^{-x_1l_1-x_2l_2}G_0(l_1,l_2;t).
\eeq

Using the CDT program one can derive the following differential equation for the Laplace transformed propagator
\beq\label{eq:CDTpropx}
\frac{\prt}{\prt t}G_0(x_1,x_2;t) = 
 \frac{\prt}{\prt x_1}\left((\lambda-x_1^2) \; G_0(x_1,x_2;t)\right).
\eeq  
In terms of the length of the spatial boundary the propagator can then also be written as
\beq\label{eq:CDTHam}
G_0(l_1,l_2;t)=\bra{l_2} e^{-t \cH_0} \ket{l_1}, \quad \cH_0 = -l \frac{\prt^2}{\prt l^2} +\lambda l.
\eeq
Here the effective quantum Hamiltonian describes the time evolution of the length of the spatial slice. In particular, one obtains that the average length of a spatial slice as well as its fluctuations are both of the same order $1/\sqrt{\lambda}$ as is also illustrated in Fig.\ \ref{fig2}. 

\section{Causal string field theory and stochastic quantization}

Viewing each spatial slice as a string one can interpret the two-dimensional fluctuating surfaces considered in the previous section as so-called worldsheets of propagating strings. In principle one can then also allow strings to split and join, thus creating surfaces of arbitrary topology. Such a so-called string field theory for CDT was introduced in \cite{cap,cdt-sft} where each splitting and joining of a string is weighted with a factor $\kappa$ according to the action \eref{eq:action}. Using this framework it was observed in \cite{matrix,topologies} that the partition function \eref{eq:partition} can then be written as the following simple integral
\beq
Z(\kappa,\lambda)= \int dx \exp\left[-\frac{1}{g_s}\left(\lambda x -\frac{1}{3}x^3 \right)\right], \quad \kappa =\frac{g_s}{\lambda^{3/2}},
\eeq
where the integral is formal in the sense that it should be understood as a suitable expansion in powers of the dimensionless string coupling $\kappa=g_s/\lambda^{3/2}$.

In view of \eref{eq:FP} and \eref{eq:eq} this suggests that the effective quantum Hamiltonian can be derived from a stochastic quantization procedure defined by the following Langevin equation \cite{stoch}
\beq
\frac{dX_t}{dt}=-\left.\frac{\partial S(x)}{\partial x}\right|_{x=X_t} +\sqrt{2 g_s} \,\nu(t),\quad S(x)=\lambda x -\frac{1}{3}x^3.
\eeq
The corresponding Fokker-Planck equation reads
\beq \label{eq:stoDiff}
\frac{\prt}{\prt t} P(x,x_0,t)=\frac{\prt}{\prt x}\left(\lambda -x^2+g_s \frac{\prt}{\prt x}  \right)P(x,x_0,t).
\eeq
One can now directly identify $P(x,x_0,t)$ with the propagator $G(x,x_0,t)$ of the string field theory. By performing an inverse Laplace transformation of \eref{eq:stoDiff} one obtains the propagator and Hamiltonian in the ``length representation''. These describe the transition amplitude of a string to propagate from a specific initial length $l_1$ to a certain final length $l_2$,
\beq\label{eq:stoHam}
G(l_1,l_2;t)=\bra{l_2} e^{-t \cH} \ket{l_1}, \quad \cH = -l \frac{\prt^2}{\prt l^2} +\lambda l -g_s l^2.
\eeq
One observes that Eqs.\ \eref{eq:stoDiff} and \eref{eq:stoHam} exactly yield \eref{eq:CDTpropx} and \eref{eq:CDTHam} for $g_s=0$. Hence we see that for $g_s=0$ we exactly obtain the CDT propagator without any spatial or space-time topology changes. Furthermore, the stochastic time $t$ which is in general a fictitious time variable exactly corresponds to the proper time on the worldsheets.

For $g_s>0$ the effective quantum Hamiltonian \eref{eq:stoHam} describes fluctuating surfaces where spatial slices are allowed to split and joint each weighted by a factor $g_s$. The corresponding amplitudes thus include a summation over all genera! It is interesting that \eref{eq:stoHam} can be used to explicitly give solutions for some amplitudes. The simplest amplitude to look at is the disc function with is a propagator with one boundary shrunken to zero and integrated over time, i.e.
\beq
W(l)=\int_0^\infty d t\, G(l,l_2=0;t)
\eeq
From \eref{eq:stoDiff} one can derive that $W(l)$ is determined by
\beq
\cH \,W(l)=0, 
\eeq
where $\cH$ is given in \eref{eq:stoHam}. This equation can be interpreted as a Wheeler-deWitt equation for the spatial slices. The explicit solution to this equation can be given in terms of Airy functions ${\rm Bi}(\cdot)$ and reads
\beq
W(l) =  \frac{{\rm Bi}\left(\frac{\lambda}{g_s^{2/3}}-g_s^{1/3} l \right)}{
{\rm Bi}\left(\frac{\lambda}{g_s^{2/3}}\right)}.
\eeq
This expression corresponds to the following genus expansion in the dimensionless string coupling $\kappa=g_s/\lambda^{3/2}$:
\bea
W(l)   & = & \raisebox{-15pt}{\includegraphics[height=40pt]{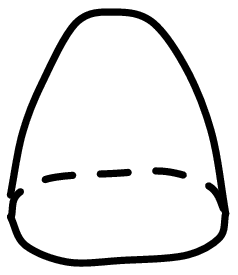}}\,+
\kappa\,\, \raisebox{-15pt}{\includegraphics[height=40pt]{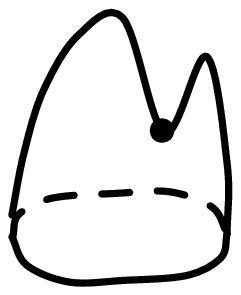}}\,+
\kappa^2 \left(\,\raisebox{-15pt}{\includegraphics[height=40pt]{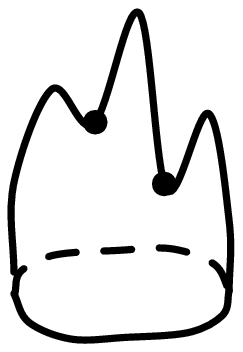}}\,+\,\raisebox{-15pt}{\includegraphics[height=40pt]{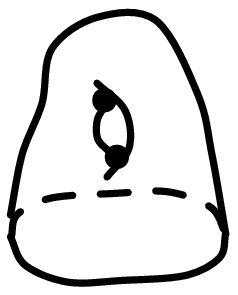}}\,\right)+ \mathcal{O}(\kappa^3).
\eea

\section{Conclusion}

We presented a recently found stochastic quantization formulation of a ``causal'' random surface model defined through CDT. This formulation is reminiscent of an analogous formulation for Euclidean quantum gravity defined through DT \cite{sto-kawai,jevicki}. The stochastic quantization procedure enabled us to determine the effective quantum Hamiltonian \eref{eq:stoHam} of the nonperturbative dynamics including the sum over topologies. A detailed physical interpretation of this Hamiltonian including a discussion on the problem of unboundedness can be found in \cite{stoch}. 

In our model the external stochastic time is identified with proper time on the string worldsheets. This special role of an external time variable and unitarity which is inherent to the CDT formulation are very similar to the setup of Ho\v{r}ava-Lifshitz gravity \cite{horava} for which the concept of stochastic quantization might also be relevant. This points towards a possible relation between these two models. In fact, it was observed very recently that CDT might not only serve as a lattice regularization of the renormalization group approach to quantum gravity, but also for Ho\v{r}ava-Lifshitz gravity \cite{cdtmeetsHL}.

\subsection*{Acknowledgments}
JA, RL, WW acknowledge support by
ENRAGE (European Network on
Random Geometry), a Marie Curie Research Training Network, 
contract MRTN-CT-2004-005616  in the 
European Community's Sixth Framework Programme.
RL acknowledges support by the Netherlands
Organisation for Scientific Research (NWO) under their VICI
program. SZ would like to thank the Department of Statistics at S\~ao Paulo University (IME-USP) as well as the Institute for Pure and Applied Mathematics (IMPA) for kind hospitality and the ISAC program (Erasmus Mundus) for financial support.

\section*{References}

\begin{thebibliography}{99}

\bibitem{al}
J.~Ambj\o rn and R.~Loll,
Nucl.\ Phys.\ B { 536} (1998) 407-434
[hep-th/9805108].

\bibitem{cap}
  J.~Ambj\o rn, R.~Loll, W.~Westra and S.~Zohren,
  JHEP {\bf 0712} (2007) 017
  [arXiv:0709.2784 [gr-qc]].
  
\bibitem{stoch}
  J.~Ambj{\o}rn, R.~Loll, W.~Westra and S.~Zohren,
  Phys.\ Lett.\  B {\bf 680}, 359 (2009)
  [arXiv:0908.4224 [hep-th]].
  
\bibitem{parisi}
  G.~Parisi and Y.~Wu,
  Sci.\ Sin.\  {\bf 24}, 483 (1981).
  
\bibitem{dam}
  P.~H.~Damgaard and H.~Huffel,
  Phys.\ Rept.\  {\bf 152}, 227 (1987).
  
\bibitem{dijk}
  R.~Dijkgraaf, D.~Orlando and S.~Reffert,
  Nucl.\ Phys.\  B {\bf 824}, 365 (2010)
  [arXiv:0903.0732 [hep-th]].
  
\bibitem{review}
  J.~Ambj{\o}rn, R.~Loll, Y.~Watabiki, W.~Westra and S.~Zohren,
  arXiv:0911.4208 [hep-th].

\bibitem{cdt-sft}
  J.~Ambj\o rn, R.~Loll, Y.~Watabiki, W.~Westra and S.~Zohren,
  JHEP {\bf 0805} (2008) 032
  [arXiv:0802.0719 [hep-th]].
  
\bibitem{matrix}
  J.~Ambj{\o}rn, R.~Loll, Y.~Watabiki, W.~Westra and S.~Zohren,
  Phys.\ Lett.\  B {\bf 665}, 252 (2008)
  [arXiv:0804.0252 [hep-th]].
  
\bibitem{topologies}
  J.~Ambj{\o}rn, R.~Loll, W.~Westra and S.~Zohren,
  Phys.\ Lett.\  B {\bf 678}, 227 (2009)
  [arXiv:0905.2108 [hep-th]].
  
\bibitem{sto-kawai}
  M.~Ikehara, N.~Ishibashi, H.~Kawai, T.~Mogami, R.~Nakayama and N.~Sasakura,
  Prog.\ Theor.\ Phys.\ Suppl.\  {\bf 118} (1995) 241
  [arXiv:hep-th/9409101].

\bibitem{jevicki}
  A.~Jevicki and J.~P.~Rodrigues,
  Nucl.\ Phys.\  B {\bf 421} (1994) 278
  [arXiv:hep-th/9312118].
  
\bibitem{horava}
 P.~Ho\v{r}ava,
  Phys.\ Rev.\  D {\bf 79} (2009) 084008
  [arXiv:0901.3775 [hep-th]].
  
\bibitem{cdtmeetsHL}
  J.~Ambj{\o}rn, A.~G\"orlich, S.~Jordan, J.~Jurkiewicz and R.~Loll,
  arXiv:1002.3298 [hep-th].
 
\end{thebibliography}

\end{document}